# Lightwave-Electronic Harmonic Frequency Mixing

Matthew Yeung,[1*] Lu-Ting Chou,[1,2] Marco Turchetti,[1] Felix Ritzkowsky,[1] Karl K. Berggren,[1] Philip D. Keathley[1*]

[1]Research Laboratory of Electronics, Massachusetts Institute of Technology, 77 Massachusetts Ave, Cambridge, MA 02139, USA
[2]Institute of Biophotonics, National Yang Ming Chiao Tung University Linong Street, Beitou District, Taipei City, 112304, TW

[*]To whom correspondence should be addressed; E-mail: myeung@mit.edu, pdkeat2@mit.edu.

**Electronic frequency mixers are fundamental building blocks of electronic systems. Harmonic frequency mixing in particular enables broadband electromagnetic signal analysis across octaves of spectrum using a single local oscillator. However, conventional harmonic frequency mixers do not operate beyond hundreds of GHz to a few THz. If extended to the petahertz scale in a compact and scalable form, harmonic mixers would enable field-resolved optical signal analysis spanning octaves of spectra in a monolithic device without the need for frequency conversion using nonlinear crystals. Here we demonstrate lightwave-electronic harmonic frequency mixing beyond 0.350 PHz using plasmonic nanoantennas. We demonstrate that the mixing process enables complete, field-resolved detection of spectral content far outside that of the local oscillator, greatly extending the range of detectable frequencies compared to conventional heterodyning techniques. Our work has important implications for applications where optical signals of interest exhibit coherent femtosecond-scale dynamics spanning multiple harmonics.**

## Introduction

Lightwave electronics (also often called PHz electronics) seek to integrate optics and electronics effectively, leveraging sub-cycle information contained within the ultrafast oscillations of light fields (Krausz and Stockman 2014; Borsch et al. 2023; Schoetz et al. 2019; Ciappina et al. 2017; Karnetzky et al. 2018). In this pursuit of electronics operating at optical frequencies, a significant obstacle arises from the mismatch between the characteristic frequencies of optical (PHz regime) and conventional electronic systems for readout (GHz-THz). To solve similar issues in frequency mismatch in more conventional radio-frequency electronics, nonlinear frequency mixers are used, with myriad applications including radar, cellular phone service, and radio communications. Harmonic frequency mixers in particular enable the use of a single local oscillator to capture information from both the fundamental as well as higher-order harmonic frequency channels, bringing their information content down to lower, baseband frequencies for readout (Evenson et al. 1972) (see Fig. 1c). Compact petahertz-electronic harmonic frequency mixers would enable field-resolved optical signal analysis spanning octaves of the optical spectrum within a single device. Here we demonstrate the use of plasmonic nanoantennas as lightwave electronic harmonic frequency mixers (see Fig. 1) for the field-resolved characterization of harmonic optical waveforms (PHz-scale).

To provide a more flexible, field-resolved readout of optical signals using electronic systems, early efforts in the 1970s aimed to extend electronic harmonic frequency mixing techniques to mid-infrared frequencies (up to 88 THz) using metal point-contact diodes (Evenson et al. 1972). However, progress stagnated until recent advancements in optical and nanofabrication technologies. Recent work has shown that nanoscale needle tips and plasmonic antennas having nanoscale vacuum channels act as nonlinear electronic diode elements similar to the earlier point-contact diodes (Evenson et al. 1972; Small et al. 1974). Through their carrier-envelope-phase sensitivity (Rybka et al. 2016; Ludwig et al. 2020; Putnam et al. 2017; Keathley et al. 2019; Yang et al. 2020; Hanus et al. 2023), use in field sampling (Bionta et al. 2021; Blöchl et al. 2022), and measurements of their photoemission response (H. Y. Kim et al. 2023; Wimmer et al. 2014; Dombi et al. 2013; Dienstbier et al. 2023; Ropers et al. 2007), researchers have demonstrated that their electronic response can extend up to 1 PHz and beyond.

While past work has demonstrated the use of sub-cycle, optical-field emission from nanoantennas and needle-tips for field-resolved optical waveform characterization (Bionta et al. 2021; Blöchl et al. 2022) these demonstrations were limited to CEP-stable few-cycle waveforms and frequencies contained within the spectrum of the gate waveform driving the electron emission. Here we experimentally demonstrate how multicycle, non-CEP-stable sources can be used for field-resolved analysis extending to signal waveforms with central frequencies well higher than that of the driving gate waveform through harmonic frequency mixing. Specifically, we show that the highly nonlinear, broadband electronic response of plasmonic nanoantennas enables lightwave-electronic harmonic frequency mixing into the petahertz regime for optical signal processing.

In our proof-of-concept measurement, we use electrically-connected nanoantenna devices for accurate amplitude and phase-resolved readout of both the fundamental (0.177 PHz, 1690 nm) and second harmonic (0.353 PHz, 850 nm) fields of an optical waveform using only the fundamental waveform as the local oscillator (see Fig. 1 for an overview). Our measurements demonstrate how under multicycle operation, petahertz-electronic nanoantennas can be conceptualized and used as harmonic frequency mixers to greatly extend the bandwidth of time-domain, field-resolved optical detection beyond one octave of spectral coverage without the need for prior nonlinear conversion in crystals, spectral phase retrieval, single-cycle waveform generation, or carrier-envelope-phase stabilization. Our study highlights a crucial connection between lightwave electronics and traditional nonlinear electronics. This connection serves to unite the electronics and optical physics communities, filling a gap in existing literature predominantly focused on strong-field and optical physics. By clarifying this link, this work acts as a bridge between these two fields.

The increased bandwidth obtained through harmonic mixing enables seamless amplitude- and phase-resolved characterization of nonlinear processes of interest, such as solid-state harmonic generation (Z. Wang et al. 2017; Keathley et al. 2023; Huang et al. 2018; Schubert et al. 2014; Hohenleutner et al. 2015; Ito et al. 2023; Jin et al. 2018; Reimann et al. 2018; Schmid et al. 2021; Langer et al. 2017), coherent Raman scattering (Marks and Boppart 2004; Kumar et al. 2018), and multiphoton processes (Alves et al. 2022; Wen et al. 2019; T. Wang and Xu 2020; Evans and Xie 2008), without the need for nonlinear frequency conversion, spectral phase retrieval, or a spectrally-overlapped local oscillator reference. In the far term, we anticipate lightwave-electronic harmonic mixer devices will provide basic building blocks for field-resolved electromagnetic signal detection and processing at optical frequencies.

## PHz Harmonic Mixing for Optical Waveform Analysis

Recent work has shown that cross-correlation using the nonlinear photoemission from gases and nanostructures as the electronic readout enables field-resolved optical waveform characterization with sub-cycle resolution (Schultze et al. 2013; Schiffrin et al. 2013; Sederberg et al. 2020; Higuchi et al. 2017; Park et al. 2018; Bionta et al. 2021; D. Zimin et al. 2021; Blöchl et al. 2022; Altwaijry et al. 2023; Hui et al. 2022; K. T. Kim et al. 2013; Herbst et al. 2022). Each technique starts with a strong gate waveform that drives sub-cycle photoelectron emission (red curve in Fig. 1a). A weak signal (blue curve in Fig. 1a) polarized along the gate and perturbs the photoemission response (perturbative method, see for *e.g.* Refs. (Park et al. 2018; Blöchl et al. 2022; Cho et al. 2019; Liu et al. 2022)) or is cross-polarized and shifts the electron momentum (streaking-like method, see for *e.g.* Refs. (Hui et al. 2022; K. T. Kim et al. 2013; Wirth et al. 2011; Borrego-Varillas, Lucchini, and Nisoli 2022)) as a function of delay $\tau$. The streaking-like method results in a time-integrated interaction over the signal field, resulting in a delay-dependent photocurrent that is proportional to the signal's vector potential. For the perturbative method, however, the delay-dependent current relates directly to the signal's electric field through an instantaneous coupling between the signal and gate waveforms analogous to the coupling of voltage waveforms in nonlinear electronic frequency mixers. Here we focus on the perturbative method.

In this section, we first briefly introduce how these perturbative, nonlinear cross-correlation measurements provide amplitude and phase information of the signal. Our treatment focuses on the field-driven photoemission response from asymmetric nanoantenna structures like those used in our experiment but could be extended to other systems. Following this introduction, we show how these perturbative cross-correlation measurements can be viewed through the lens of nonlinear electronic frequency mixing. This framing allows us to better understand how nonlinear, field-driven photoelectron emission devices provide field-resolved readout across spectral harmonics using only a single local oscillator without the need for carrier-envelope-phase stabilization. It also provides a framework for understanding how we might translate technologies used now at lower frequencies (*e.g.* RF or microwave) into the petahertz regime.

When driven by intense, few-cycle waveforms polarized parallel to the tip axis, sub-optical-cycle tunneling current (Rybka et al. 2016; Ludwig et al. 2020) can be emitted from a nanoantenna (connected triangle features in Fig. 1a,b) to a collector (adjacent wires in Fig. 1a,b). This configuration leads to a half-wave rectified current response relative to the driving field at the antenna apex. In Fig. 2a we show a calculation of the sub-cycle electron emission (blue) as a function of a few-cycle gate field (red) for two different values of carrier-envelope phase (solid versus dashed lines). Note that we have used shorter gate pulse durations in Fig. 2 for illustration purposes. The current emission was modeled using the Fowler-Nordheim (FN) rate equation as described in Refs. (Putnam et al. 2017; Keathley et al. 2019; Yang et al. 2020; Bionta et al. 2021; Ritzkowsky et al. 2023; Blöchl et al. 2022) (see further discussion in supplementary section 2.3.1).

A nonlinear cross-correlation measurement similar to that discussed in Refs. (Park et al. 2018; Blöchl et al. 2022; Altwaijry et al. 2023) can be modeled by using a strong gate waveform which in addition to a weak signal waveform (to be measured). When the signal and gate are superimposed with the same polarization and the relative delay $\tau$ between the two is varied, the time-averaged charge transferred across the nanoantenna gap can be modeled as

$$I(\tau) = q \int_{-T_{rep}/2}^{T_{rep}/2} \Gamma\left[E_{\text{gate}}(t-\tau) + E_{\text{signal}}(t)\right] dt$$

where $\Gamma$ is the FN equation. Provided that $E_{\text{signal}}$ is sufficiently small, such that the electric field perturbation seen by the nanoantenna can be assumed to be linear and $E_{\text{gate}}$ can be Taylor-expanded to the first order. The resulting expression is given by

$$I(\tau) \approx q \int_{-T_{rep}/2}^{T_{rep}/2} \Gamma\left[E_{\text{gate}}(t-\tau)\right] + \left[\frac{d\Gamma}{dE}|_{E_{\text{gate}}(t-\tau)} \cdot E_{\text{signal}}(t)\right] dt$$

In this equation, the integral of the second term corresponds to the measured electric field waveform. This integral represents the small-signal cross-correlation between $\frac{d\Gamma}{dE}|_{E_{\text{gate}}(t-\tau)}$ and $E_{\text{signal}}(t)$ which we denote as $I_{cc}(\tau)$. Since this expression represents a cross-correlation, in general, the Fourier-transformed expression can be written as $\tilde{I}_{cc}(\omega) \approx \mathcal{F}\left[\frac{d\Gamma}{dE}|_{E_{\text{gate}}}\right]^* \cdot \tilde{E}_{\text{signal}}(\omega)$ where $\tilde{H}(\omega) = \tilde{I}_{cc}(\omega)/\tilde{E}_{\text{signal}}(\omega) = \mathcal{F}\left[\frac{d\Gamma}{dE}|_{E_{\text{gate}}}\right]^*$ is the full complex frequency response of the output relative to the signal field, which is plotted as a function of in Fig. 2c.

Note that in Fig. 2c we see that the detector frequency response contains frequency components at harmonic orders both higher and lower than that of the central frequency of the gate. The operating principle behind this is identical to that of a nonlinear electronic harmonic frequency mixer where the gate generates the local oscillator with $f_{\text{LO}} = f_{\text{gate}}$, $f_{\text{LO}}$ being the local oscillator frequency and $f_{\text{gate}}$ the central frequency of the gate waveform. Due to the high nonlinearity of the FN tunneling rate $\Gamma$ and half-wave rectification, the current response contains frequencies outside of the optical local oscillator centered at every integer harmonic (Bionta et al. 2021). This is visualized in Fig. 2b where we show how the sub-cycle burst in charge over one period can be expressed as a sum of harmonic frequency components. These electronic frequency components effectively serve as frequency-distributed local oscillators that mix with the small signal. The difference frequency components then provide the baseband response for amplitude and phase-resolved readout.

Conceptualizing the devices as electronic optical frequency mixers aids in describing important properties of the devices. First, it becomes apparent that carrier-envelope phase (CEP) locking of the gate and signal pulse is not a requirement for amplitude and phase-resolved waveform readout even for the case of signals comprised of higher-order harmonics provided that the gate and signal exhibit relative phase locking. Consider the case of a perfectly sinusoidal gate (*i.e.* local oscillator) and signal functions, where the signal is a harmonic of the gate. Let the signal then be a harmonic of the gate frequency $f_{\text{sig}} = k f_{\text{gate}}$ as represented in Fig. 1c. We can then represent $\frac{d\Gamma}{dE}|_{E_{\text{gate}}(t-\tau)}$ as an expanded series of harmonics of the gate frequency

$$\frac{d\Gamma}{dE}|_{E_{\text{gate}}(t-\tau)} = h_0 + \frac{1}{2}\left(\sum_{n=1}^{\infty} \tilde{h}_n\, e^{in\varphi} e^{i2\pi n f_{\text{gate}}(t-\tau)} + \text{c.c.}\right)$$

where $\varphi$ represents the absolute phase shift of the gate, analogous to the CEP for the case of a pulsed gate. Likewise, the signal is represented by

$$E_{\text{signal}}(t) = \frac{1}{2}\tilde{a}_k e^{ik\varphi + \Delta\varphi} e^{i2\pi k f_{\text{gate}} t} + \text{c.c.}$$

where $\Delta\varphi$ represents any remaining phase difference in addition to $k\varphi$. The DC output response is then formed by the multiplication of conjugate and non-conjugate coefficients of $\tilde{h}_k$ and $\tilde{a}_k$ and is found to be

$$I_{\text{cc}}(\tau) = \frac{1}{4}\tilde{h}_k^* \tilde{a}_k e^{i2\pi k f_{\text{gate}} \tau} e^{i\Delta\varphi} + \text{c.c.}$$

Note that since the signal and current responses are both phase-locked to the local oscillator (gate), the absolute phase terms $\varphi$ always cancel and the harmonic mixing response is not sensitive to any fluctuations of the absolute phase $\varphi$.

These behaviors translate directly to the case of pulsed gate and signal inputs. A finite envelope of the gate pulse leads to a broadening of the harmonic pass bands as shown in Fig. 2c. These shifts are equal and opposite to those of optically-generated harmonic pulses meaning that so long as the source of harmonics and the nonlinear optical radiation exhibit relative phase locking, the sampled response does not depend on the absolute carrier-envelope phase. We note here that absolute CEP information is effectively lost for a multi-cycle gate and a CEP-unstable laser. However, using a near single-cycle gate with a CEP-stable laser would allow one to extend this field-resolved measurement to a sampling measurement where the true electric field is measured, rather than a measurement of an ensemble of CEP values.

In addition to highlighting the role of CEP dependence on the readout, we see the importance of high nonlinearities and rectification in extending device bandwidth across harmonics. While one can certainly obtain a mixing response via conventional heterodyning and homodyning using an $E^2$ detector (*e.g.* a photodiode) (Ferdous et al. 2009), we see from this analysis that an $E^2$ current response without rectification would not yield higher harmonic components in the electronic response. But higher harmonic terms appear at every integer harmonic for higher-order nonlinearities and half-wave rectification. For the plasmonic detectors in this work, these nonlinearities can exceed $E^{10}$.

# Results and Discussion

## Experimental Demonstration

In this section, we present experimental results demonstrating PHz harmonic mixing through field-resolved optical waveform characterization. First, we performed degenerate waveform characterization where we used a 10-cycle near-infrared gate and signal waveforms both having a central frequency of 0.177 PHz, corresponding to when $k = 1$ in Eqn. [eqn:signal]. This measurement illustrates the capability of multicycle sampling without the need for CEP stabilization. We next present the characterization of the second harmonic waveform (0.353 PHz) using the fundamental (10-cycle 0.177 PHz) as the gate to demonstrate harmonic frequency mixing, corresponding to $k = 2$ in Eqn. [eqn:signal].

In our measurements, we used an asymmetric nanoantenna design as shown in Fig. 3a as this naturally breaks inversion symmetry and enables frequency response ($\tilde{H}(\omega)$) to have both even and odd integer harmonics (Ritzkowsky et al. 2022). Gold was chosen as the antenna material and fused silica as the substrate. In Fig. 3b we plot the frequency-dependent field enhancement (black curve) and group delay (red curve) at the antenna apex relative to the incident light for our chosen design. The spectral regions covered by the fundamental (shaded red) and its second harmonic (shaded blue) are also shown for reference. The nominal nanoantenna geometry was chosen to have a resonant frequency between that of the fundamental and second harmonic, corresponding to a triangle base width of 180 nm and a height of 240 nm. This choice was for two reasons. First, it results in a compromise where the fundamental and second harmonic both have maximal field enhancement ($\approx 15 \times$ and $\approx 10 \times$, respectively). Second, since the fundamental and second harmonic both excite the antennas off-resonance, they experience a negligible amount of intensity reshaping and group delay dispersion (less than 2.5 fs change in group delay over the bandwidth of the fundamental and second harmonic), meaning that the excited field waveforms at the antenna apex are not significantly reshaped in time relative to the incident field waveforms. Furthermore, this demonstrates that operating the devices on-resonance is not necessary for the practical use of optical waveform analysis. For more details regarding device design and fabrication see Supplemental Section 3.

To perform the degenerate waveform characterization, we used a 10-cycle (57 fs) 0.177 PHz (1690 nm) pulse from a commercial optical parametric amplifier (LightConversion Cronus 3P) for both the gate and signal. This measurement differs from the work of Bionta et al. (Bionta et al. 2021) which uses the on-resonance property of the nanoantennas to perform degenerate waveform characterization using a 2.5-cycle pulse. The gate and signal pulses were prepared using a Mach–Zehnder interferometer (MZI), and the pulse duration was confirmed through frequency-resolved optical gating (FROG) measurements (See Fig. 9 for the measurement schematic, Fig. 10 for FROG (Section 5.2).

Before performing the cross-correlation measurements, we first confirmed that the devices were operated in the optical-field emission regime through the analysis of current scaling with intensity by measuring the output current as a function of the incident gate pulse energy and verifying that the emission could be well-described through a quasi-static Fowler-Nordheim tunneling rate (see Fig. 8 and supplementary section 2.3.1 for further details). Having confirmed operation in the optical-field emission regime, we then illuminated the antennas with a gate pulse with an energy of 5.4 nJ (0.85 V/nm) and a small signal with an energy of 4.4 pJ (24 V/$\mu$m) to obtain $I_{cc}(\tau)$ (red curve in Fig. 4). In the experiment, the signal arm was chopped and measured using lock-in detection to isolate $I_{cc}(\tau)$ from the constant background current. The small signal gain from the high nonlinearity of the optical-field emission response enabled the sampling of signal pulse energies on the order of 1000× smaller than those of the gate. The measured optical period was 5.6 fs which matched the expected value for a frequency of 0.177 PHz. The pulse of the sampled field was 57 fs full-width at half maximum (FWHM), in close agreement with the 58 fs FWHM pulse duration retrieved FROG measurement (see Fig. 10c).

Spectral intensity and group delay analysis are shown in Fig. 4b. We find good agreement between $|\tilde{I}_{cc}(\omega)|^2$ (solid red line) and the spectral intensity as measured using a commercial grating-based indium gallium arsenide spectrometer (solid black line) after normalization. Also, the extracted group delay of $\tilde{I}_{cc}(\omega)$ (dashed red line) is shown to be concave up and agrees well with the group delay retrieved from the FROG measurement (dashed black line). These results further confirm that $I_{cc}$ accurately represents the signal field of the fundamental to within a constant phase offset.

We then performed the non-degenerate waveform characterization of the second harmonic signal waveform via harmonic mixing using the same 10-cycle gate as before ($f_{gate}$ = 0.177 PHz) with the signal being the second harmonic of the gate waveform ($f_{sig}$ = 0.353 PHz). We note that this measurement is different from other studies using both the first and second harmonic ($\omega - 2\omega$). The work of Arai et al. (Arai et al. 2023) utilizes $\omega - 2\omega$ for optical phase measurements, whereas Dienstbier et al. (Dienstbier et al. 2023) use the $\omega - 2\omega$ to break symmetry for the application of ultrafast switching. In this measurement, we use a sufficiently weak second harmonic waveform to demonstrate field-resolved waveform characterization.

The second harmonic was generated by adding lenses and a type 1 beta-barium borate ($\beta$-BaB2O4, BBO) crystal to the signal arm of the MZI (see Fig. 12 for measurement schematic). We illuminated the antennas on the device using a pulse energy of 4.9 nJ (0.80 V/nm) for the gate and a pulse energy of 23 pJ (49 V/$\mu$m) for the second harmonic signal (after removal of the residual fundamental through filtering). The measurement of $I_{cc}(\tau)$ using the second harmonic signal is shown in Fig. 4a (blue curve). The measured optical period was 2.8 fs matching the expected optical period for a center frequency of 0.353 PHz.

To verify the accuracy of the sampled field, we again performed FROG and used a commercial grating-based silicon spectrometer as a reference in the same way as in the degenerate case. The pulse duration from our FROG measurement was found to be 49 fs FWHM which matches almost perfectly with the duration of $I_{cc}(\tau)$, being 48 fs FWHM (see supplementary information Fig. 13 for a comparison of FROG and $I_{cc}(\tau)$). In Fig. 4c we find that $\left|\tilde{I}_{cc}(\omega)\right|^2$ (solid blue line) is in general agreement with the spectral intensity measured using the silicon spectrometer (solid black line) after normalization. There is only a discrepancy in the intensity of the spectral wings. However, the group delay of the $\tilde{I}_{cc}(\omega)$ (solid orange curve) is concave down and matches relatively well with the FROG-retrieved group delay (dashed orange curve). Together, these results lead us to conclude that $I_{cc}(\tau)$ accurately represents the second harmonic signal's field waveform to within a constant phase offset.

Several aspects of this measurement are important to emphasize. First, given that the second harmonic generation used to generate the signal was phase-locked to the second harmonic of the sampling response as noted in Section 1, there was no need for CEP stabilization of the gate. Second, the nonlinearity of the mixing process enabled the phase-resolved, interferometric pulse readout without the need for nonlinear optical generation of a spectrally-overlapped local oscillator as would typically be required for all-optical homodyning or heterodyning techniques using a standard $E^2$ detector (D. A. Zimin, Yakovlev, and Karpowicz 2022; Ferdous et al. 2009; Mamaikin et al. 2022; Ridente et al. 2022; Keiber et al. 2016). Third is that the direct cross-correlation outputs accurately represented the fundamental and signal waveforms as confirmed using FROG without the need for any post-analysis, such as phase retrieval, or any change in the detector setup aside from the change in the signal waveform. Finally, we emphasize that for the case of the second harmonic measurement, the signal pulse was appreciably shorter than the gate (48 fs FWHM signal duration versus 60 fs FWHM gate pulse duration), and of a significantly higher carrier frequency. While techniques such as electro-optic sampling can also provide field information from non-spectrally overlapped signals, the signal frequency must be lower than that of the gate, with gate pulse envelopes shorter than the cycle time of the signal (Sulzer et al. 2020; Pupeza et al. 2020; Steinleitner et al. 2022; Lesko et al. 2021).

## Concluding Remarks and Outlook

In this study, we used non-resonant nanoantenna networks to demonstrate a broadband, on-chip electronic optical frequency harmonic mixer using optical-field-driven tunneling. We showed how the harmonic frequency mixing process enables accurate field-resolved readout of optical signal waveforms spanning more than one octave of bandwidth using a commercial laser without the need for single-cycle pulse generation or carrier-envelope phase locking. If these devices were produced at scale, calibrations similar to those required for commercial spectrometers would be necessary. These calibrations, based on electromagnetic simulations, would ensure accurate field-resolved measurement across the entire device bandwidth.

In comparison to wave mixing in crystals, the optical-field-driven tunneling mechanism provides access to higher-order nonlinearities, and thus larger mixing bandwidths, while eliminating the need for phase-matching or a separate photodetection element (D. A. Zimin, Yakovlev, and Karpowicz 2022; Fresch et al. 2023). Furthermore, unlike FROG and other spectral characterization methods, the measurements provided direct amplitude and phase information in the time domain and did not require broadband spectral measurements or phase-retrieval algorithms. While techniques such as electro-optic sampling can provide similar time-domain information, they require CEP-stabilization and gate pulse envelopes shorter than the cycle time of the signal (Sulzer et al. 2020).

While we demonstrated harmonic frequency mixing through pulse characterization, we believe similar devices will be used to create compact and sensitive optical oscilloscopes with bandwidths spanning multiple octaves. We anticipate that such optical field oscilloscopes will provide needed time-domain detection tools that will enable new approaches to the field-resolved investigation of ultrafast light-matter interactions and help accelerate the development of ultrafast source technologies (e.g. compact frequency combs and optical waveform synthesizers). Beyond time-domain, field-resolved detection, petahertz-electronic mixers could also be incorporated as fundamental components within future lightwave electronic systems for PHz-scale communication and computation. Aside from conversion to baseband, they could be used to generate new sum and difference frequency signals to be routed to other nearby on-chip devices.

## Materials and Methods

### Nanofabrication

We started with 1 cm × 1 cm fused silica pieces (MTI Corp.) and cleaned them using piranha for 10 minutes before use. For the nanoantenna array fabrication, we spin-coated polymethyl methacrylate A2 (Microchem) at 2,750 revolutions per minute and baked at 180 °C for 2 minutes. Afterward, DisCharge H2O X2 (DisChem Inc.) was spun at 3,000 revolutions per minute so that charging did not occur during the electron beam lithography write. The electron beam lithography was performed at 125 keV with a dose ranging from 4000-6000 $\mu C/cm^2$ with proximity effect correction. Development of the exposed polymethyl methacrylate samples was done at 0 °C in a solution of 3:1 2-propanol to methyl isobutyl ketone for 50 seconds. Electron beam evaporation was performed at $2 \times 10^{-6}$ Torr where we first deposited a 2 nm titanium adhesion layer, then 20 nm of gold. Lift-off was performed by submerging the samples in a 65 °C solution of N-methyl pyrrolidone (Microchem) for one hour.

Contacts were made to the nanoantenna using photolithography. We spin coat nLOF 2035 at 3,000 revolutions per minute, then bake the resist at 110 °C for 90 seconds. The exposure was performed using a maskless aligner with a wavelength of 375 nm and at a dose of 300 300 mJ/cm². After exposure, we do a post-exposure bake at 110 °C for 90 seconds, then develop for 90 seconds in AZ726. We then use electron beam lithography to deposit 10 nm of a titanium adhesion layer and 50 nm of gold. Liftoff was performed at room temperature in a solution of acetone for at least 6 hours. The samples were ashed for 30 seconds, then mounted on a printed circuit board (PCB), and wire bonded.

## Laser and measurement Methods

We used a LightConversion optical parametric amplifier pumped by a Yb:KGW laser pulse picked to 1 MHz (Cronus 3P) for our experiments. The idler output was compressed using a prism pair. The samples on the PCB were used as-is in ambient conditions and the output was connected to a trans-impedance amplifier with a gain of 1 V/nA (FEMTO). We measured the signal pulse-induced current ($\Delta I$) through a lock-in measurement referenced by chopping the signal arm at ~ 277 Hz. The x- and y-channels of the lock-in are output to an oscilloscope (Keysight) with a sampling rate of at least 25-50 kSa/s. Before sampling measurements, we illuminated the device with the gate pulse and ensured that the photocurrent remained constant for at least several minutes. To temporally control the delay between the signal and gate pulse, we placed the gate pulse on a closed-loop piezo stage (Piezosystem Jena) with $\pm$ 14 nm (0.05 fs) repeatability. The scan time for each trace was set to 20 seconds and averaged accordingly. We continuously scan the delay stage in a certain direction with constant velocity to reduce measurement variation.

In every measurement, we used neutral density (ND) filters to control the power of the signal and gate pulses. After recombining the two pulses, we placed a linear film polarizer with an extinction ratio of $\geq 10^5$ (across all wavelengths) to ensure the two beams were horizontally polarized. Once collinearity was ensured between the gate and signal arm, both the signal and gate pulses were focused onto the nanoantenna chip through a reflective objective (Ealing). See Fig. 9 for the schematic.

For frequency doubling, we used a type 1, 1.5 mm thick beta-barium borate ($\beta$-BaB$_2$O$_4$, BBO) crystal with $\theta$=24 $\phi$=90 to generate the second harmonic (SHG) of $f_{LO}$ = 0.177 PHz. The SHG (0.353 PHz) is filtered using a 0.207 PHz (1450 nm) high-frequency pass filter with ND 2 at 0.177 PHz along with a broadband achromatic half waveplate used to rotate the SHG polarization from vertical to horizontal, to match the gate pulse. Note that the integration time was crucial for accurate measurement of the higher frequencies.

For the non-degenerate supercontinuum measurement using a 10-cycle $f_{LO}$ = 0.177 PHz, we ensured that the phase induced by the optics was equivalent to when the degenerately sampled supercontinuum was measured. We also utilize the same device for degenerate and non-degenerate sampling.

For frequency-resolved optical gating (FROG) measurements, we used the RANA approach (see Ref. (Jafari, Jones, and Trebino 2019)) as a robust retrieval algorithm. The MATLAB code is available on the Trebino Group website. We utilized a 128x128 grid for the FROG reconstruction.

We padded the retrieved waveforms with zeros before taking an FFT to improve spectral resolution. This was justified as the spacing between consecutive pulses in time in the experiment is much larger (one microsecond) than the time window of the retrieved waveforms.

To calculate the bare field strengths, we took the pulse energy and converted it to field strength using the corresponding intensity and a Gaussian beam approximation in time/space. For the fundamental beam, the beam spot size corresponds to 12.8 $\mu$m × 9.54 μm at the focus (FWHM). For the second harmonic, the beam spot size corresponds to 12.1 μm × 19.1 μm at the focus.

## Electromagnetic, Sampling, and Supercontinuum Generation Simulations

For the field enhancement simulations, we used open-source finite-difference time-domain package (FDTD) PyMeep (Oskooi et al. 2010). For the gold nanoantenna and silicon oxide substrate, we used the standard materials library included in the Python package. For setting up the FDTD conditions, we used periodic boundary conditions in the nanoantenna on a silicon oxide plane with perfectly matched layers in the direction of propagation of the plane-wave source to prevent multi-reflections affecting the simulation. To obtain realistic field enhancements, we ensured that the apex of the triangle had a radius of curvature of 10 nm, which we extracted through scanning electron microscopy of our devices.

In our sampling simulations, the Fowler-Nordheim equation, expressed as $\Gamma = \alpha\phi^{-1}E^2\exp(-\beta\phi^{3/2}/E)$, was utilized. Here, $\alpha$ is $1.54\times 10^{-6}$ A eV V$^{-2}$, $\beta$ is 6.83 eV$^{-3/2}$V nm$^{-1}$, $\phi$ represents the work function (taken as 5.1 eV), and E denotes the electric field. With this equation, we numerically calculated the current cross-correlation by

$$I_{CC}(\tau) \propto \int \Gamma\left(E_{\text{gate}}(t-\tau) + E_{\text{signal}}(t)\right) dt$$

using a ratio of signal to the gate of 0.03.

## Acknowledgments

We graciously thank Prof. Sixian You for the use of the LightConversion Cronus 3P system used in this work. This work was carried out in part through the use of MIT.nano. We also graciously thank LightConversion for providing the 6-Watt Carbide used for testing degenerate sampling of a 63-cycle pulse. We would like to thank Matteo Castellani and Adina Bechofer for their comments and suggestions on the manuscript before submission. We also thank Prof. Shih-Hsuan Chia for the discussions regarding this work.

### Funding:

This work, in particular the experimental measurements and initial analysis, was supported by the Air Force Office of Scientific Research under award number FA9550-18-1-0436. Later phases of this work, in particular regarding the analytical treatment of harmonic frequency mixing and its use for field-resolved detection of harmonic fields, was supported by the U.S. Department of Energy, Office of Science, Office of Basic Energy Sciences, under Award Number DE-SC0024173. M. Y. Acknowledges support from the National Science Foundation Graduate Research Fellowship Program, Grant No. 1745302. This material is based upon work supported by the National Science Foundation MPS-Ascend Postdoctoral Research Fellowship under Grant No. 2402151. L.-T. C acknowledges financial support from the Ministry of Education (R.O.C) for the Overseas Internship Program directed by Prof. Shih-Hsuan Chia at National Yang Ming Chiao Tung University

## Supplementary Material

Supplementary Text
Supplementary Experimental Information
Figs. 5 to 15
References *(43-46)*
Data and code

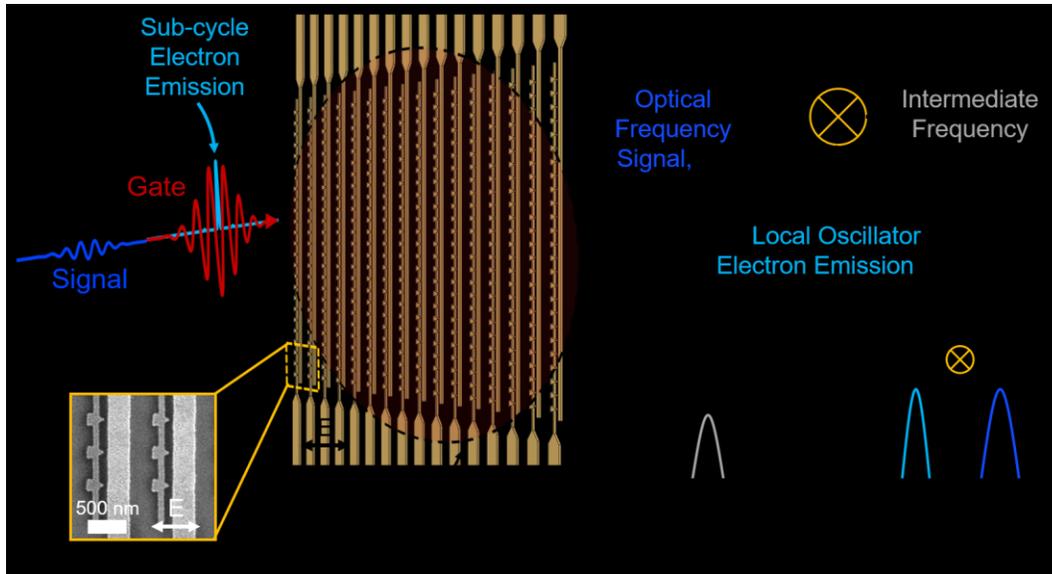

*Figure 1: **Schematic of Experimental Measurement.** (a) A gate pulse illuminates the nanoantenna network and drives sub-optical cycle electron emission. A small signal is introduced over a variable delay. This small signal modulates the electron emission from the nanoantennas leading to the optical-frequency mixing process. Both the gate and signal pulse polarization are parallel to the nanoantenna tip axis as indicated by the black arrow. (b) A representative scanning electron microscope image showing the nanoantennas and the polarization incident on the nanoantennas. (c) The devices can be conceptualized as electronic harmonic frequency mixers (top schematic) with the gate waveform of central frequency $f_{gate}$ serving as the local oscillator (LO, with central frequency $f_{LO} = f_{gate}$), and the signal having central frequency $f_{signal}$ as the optical frequency input (OF, with central frequency $f_{OF} = f_{signal}$). The mixing process (bottom schematic) provides a current signal at baseband (intermediate frequency, IF) for detection of harmonics of the local oscillator $kf_{LO}$ (right plot). Here we measure the baseband response for field-resolved measurement of the signal as a function of delay τ.*

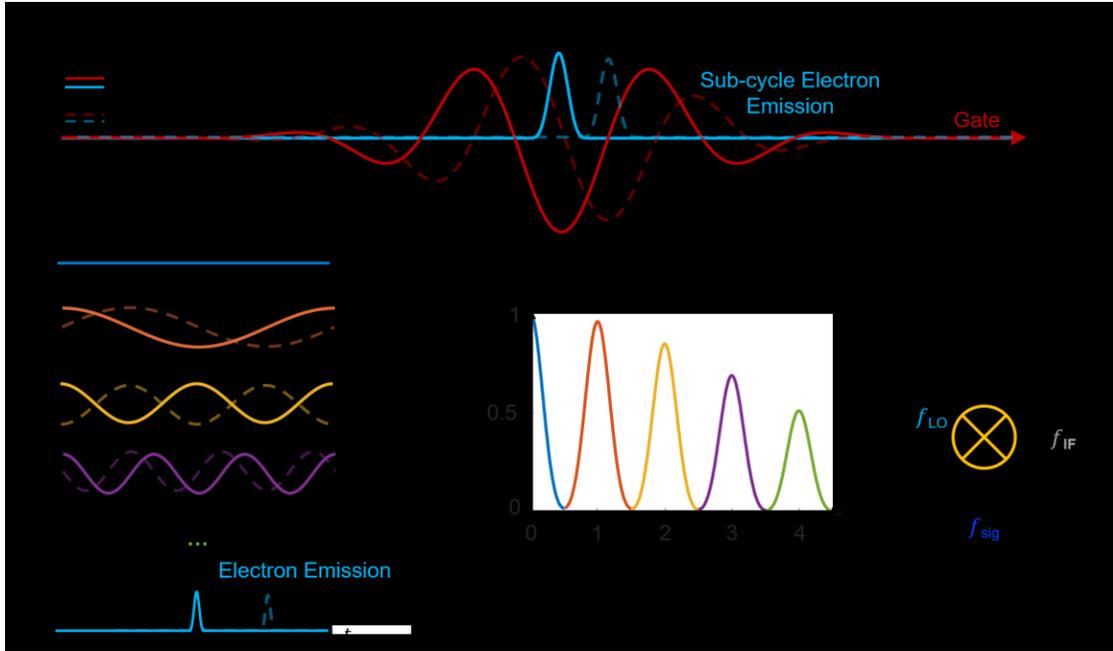

*Figure 2:  **How Sub-Cycle Emission Enables Harmonic Mixing.** (a) Depiction of sub-cycle electron emission calculated using the FN tunneling rate (teal) driven by a single-cycle pulse for a CEP = π and CEP = π/2 (dashed). (b) The sub-cycle electron emission comprises integer harmonic frequencies of the gate frequency $f_{gate}$, collectively contributing to the sub-cycle electron emission. At each of these frequencies, a phase shift occurs when the CEP of $f_{gate}$ is altered by Δϕ (here π/2). Specifically, for the fundamental frequency $f_{gate}$, the phase shift corresponds to Δϕ, while the second harmonic corresponds to 2 × Δϕ, the third harmonic to 3 × Δϕ, and subsequent higher harmonics to k × Δϕ, k is the harmonic order. (c) The calculated transfer function amplitude $|\tilde{H}(f)|$ for a 4-cycle Gaussian pulse with a center frequency of 0.177 PHz.*

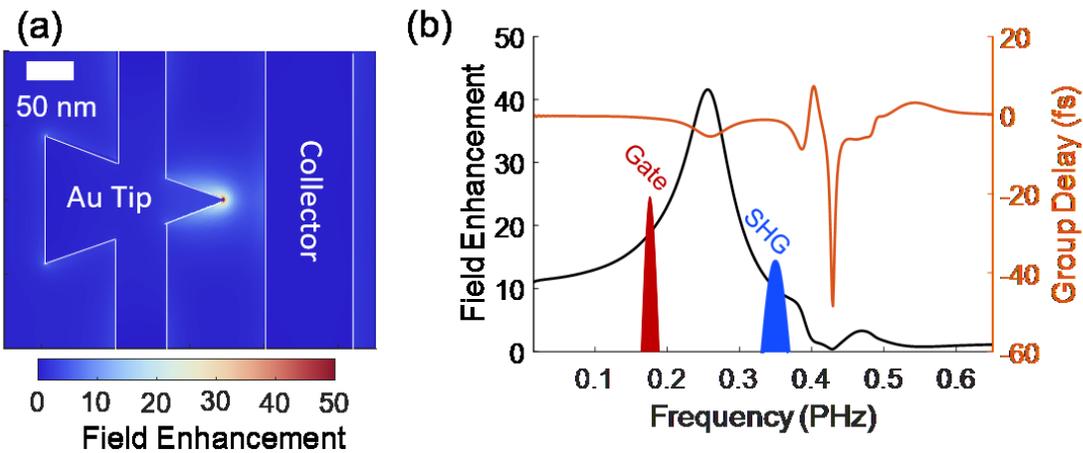

*Figure 3:  **Nanoantenna Design.** (a) Finite-difference-time-domain (FDTD) simulation of the electric field enhancement at the tip of a gold nanoantenna. (b) FDTD simulation of the field enhancement and group delay imparted by the antenna response as a function of frequency.*

*Within the spectrum, we highlight the experimental frequencies used with the gate at frequency 0.177 PHz and a higher frequency signal at 0.353 PHz, which corresponds to the second harmonic of the gate (SHG).*

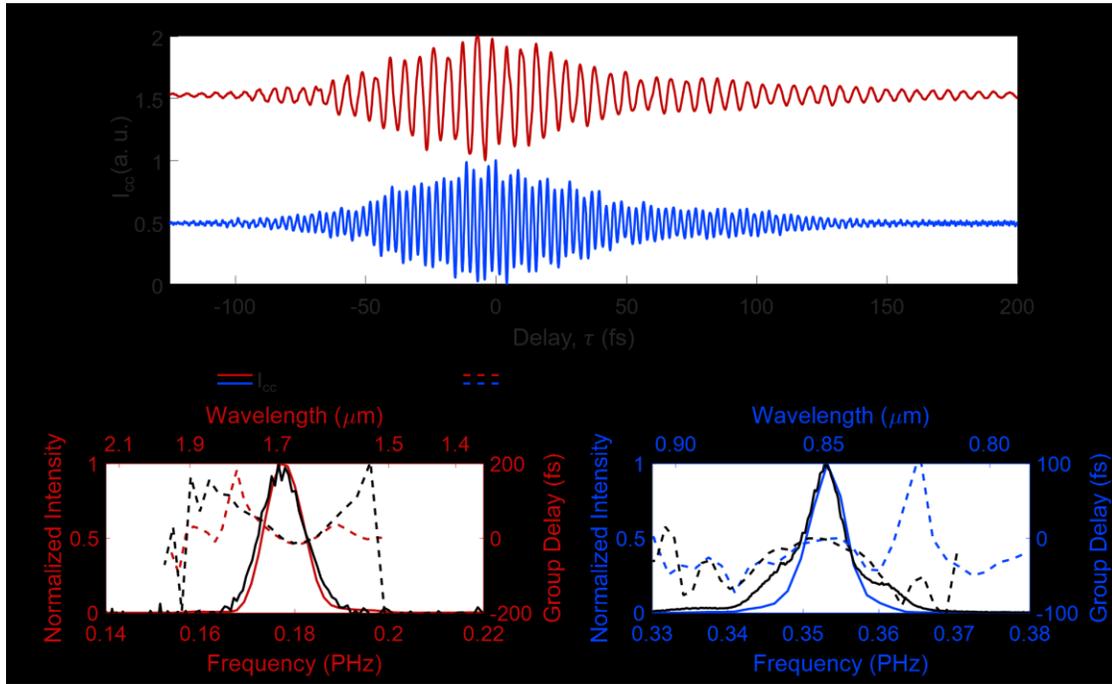

*Figure 4: **Harmonic Frequency Mixing for Field-Resolved Characterization.** (a) The measured electric field of the 10-cycle, 0.177 PHz fundamental (top, red) and measured electric field of the 0.353 PHz second harmonic (bottom, blue). The same 10-cycle, 0.177 PHz pulse was used as the gate (local oscillator) for both cases. (b) The corresponding frequency-domain intensity of the measured electric field (solid red line) compared to a commercial spectrometer (black solid line). The extracted group delay from the measured field (dashed red line) is compared to the group delay retrieved from a FROG measurement (dashed black line). (c) The corresponding frequency-domain intensity of the measured second harmonic electric field (solid blue line) compared to a commercial spectrometer (black solid line). The extracted group delay from the sampled optical field (dashed blue line) is compared to the group delay retrieved from a FROG measurement (dashed black line).*

## Supplementary materials

Supplementary Text
Supplementary Experimental Information
Figs. 5 to 15
References *(43-46)*
Data and code

# Supplementary Text

Shown in Fig. 5a are two 10-cycle Gaussian pulses used as $f_{\text{gate}}$, one with a carrier frequency corresponding to 0.353 PHz and another with 0.177 PHz. The corresponding Fowler-Nordheim electron emission rate for the two Gaussian pulses is shown in Fig. 5b. Since there are several cycles, there are several bursts of electrons that are generated. The inset displays the main burst at t=0, which is modulated by the signal pulse, enabling small signal sampling. The current burst full-width at half maximum (FWHM) determines the time resolution and is 0.28 fs for $f_{gate}$ = 0.353 PHz and 0.55 fs for $f_{gate}$ = 0.177 PHz. Lastly, the frequency response given the 10-cycle $f_{\text{gate}}$ is shown in Fig. 5c. Since we are using a 10-cycle pulse, there are multiple electron bursts in time, resulting in a reduction in the frequency bandwidth response. If we were using a single-cycle gate pulse, the frequency domain response would be the envelope shown in the black dashed lines.

Shown in Fig. 6a is the gate-dependent electron emission in time. The electron emission can be seen as the multiplication of a comb in time with the peaks separated depending on the device geometry. With an asymmetric design, the spacing is the optical period of the gate frequency. Mathematically, it can be described as the following:

$$h(t) = f(t) \cdot g(t)$$
$$k(t) = h(t) \times i(t)$$

In time, an electronic comb (f) is generated and the amplitude of the comb follows an envelope function (g), resulting in h and is purely dependent on the gate-dependent localized electric fields at the tip. Then, h is convolved with the electron emission rate envelope with finite bandwidth (i) which is directly related to the optical period of the gate frequency.

Shown in Fig. 6b is the electron emission rate for 3 different field enhancements. When the field enhancement is small, the main electron emission burst centered at t = 0 plays the largest role in perturbative sampling. As the field enhancement increases by a factor of 10 (equivalently an intensity enhancement of 1000), the side electron emission bursts start to increase in amplitude and the smaller amplitude sub-optical-cycle oscillations in the electric field start to contribute to electron emission.

## Phase information from sampling

In phase retrieval techniques like FROG, the sign of the phase is not determined. With harmonic frequency mixing, we have relative phase information. When we measure our time-domain trace, we directly take our data and FFT into the frequency domain and take the derivative of the phase. To demonstrate that we have the sign of the phase, we perform sampling simulations using a 1690 nm 2-cycle pulse as the signal and gate. The gate pulse has 0 fs² GDD and we chirp the signal pulse with + and - 200 fs² GDD. The corresponding field we measure in time is different depending on the GDD in the pulse Fig. 7a. In the frequency domain, the sign of the GD is resolved as shown in Fig. 7b.

# Supplementary Experimental Information

In this section we provide supplementary experimental information important to our work.

## Electron Emission Regime

There are numerous papers demonstrating that such nanoantenna devices operate in the optical field emission regime using 10s to 100s of pJ pulse energies with few-cycle pulses (Rybka et al. 2016; Ludwig et al. 2020; Putnam et al. 2017; Keathley et al. 2019; Bionta et al. 2021). To confirm our devices are operating in the Fowler-Nordheim regime with the 10-cycle $f_{LO}$ = 0.177 PHz gate pulse, we measured the current output versus peak field shown in Fig. 8 with a typical device that has never been used for waveform measurements, denoted as "New Device", and a device which we used for more than 4 hours of illumination (denoted as "Burned-in device"). We fit the low-pulse-energy section of the new device current vs peak field curve to the conventional multiphoton photoemission form, $I_{MP} \propto |F_0|^{2N}$, where $F_0$ is the incident peak field, finding a fit of N = 7. However, we note that there is a sharp turnover of the emission which is not fully understood.

While space charge might cause such a sharp turn-over in the emission rate, we ruled this out after considering the emitted charge densities. Considering the close to 100 pA currents and the repetition rate of 1 MHz distributed over 306 antennas, we were generating 2 electrons per pulse per antenna. Given the length of the optical pulse of approximately 10 cycles, the electrons were very unlikely to interact within the same optical half-cycle. Furthermore, from Coulomb's law, we can calculate typical forces that would result in an electron 5 nm away from the tip in an electric field strength of ≈ 0.06 V/nm, substantially lower than local fields of around 14 V $^{-1}$.

An alternative explanation is that this is emission from a surface state due to an adsorbed molecule. This is consistent with our observation that after some time of operation, we observe that the needed energy for similar emission rates increases indicated by the shift to the right of the "Burned-in Device" (red dots in Fig. 8). Our interpretation is that during the process of using the devices, surface adsorbates are removed and the emission approaches expected from Fowler-Nordheim tunneling from gold as noted in past works. In studying similar devices using field-emission from the application of DC biases, it was shown that similar behavior is due largely to water adsorption on the surface that is removed through UV radiation (Bhattacharya et al. 2023). Here we feel this removal is due to the optical illumination and interaction at the tip from the gate.

We fitted the Fowler-Nordheim curve to the "Burned-in Device" data (green curve in Fig. 8) to demonstrate that the emission is indeed well-described by quasi-static tunneling. We used the initial field of the optical waveform before enhancement in comparison with the field from this fit to determine the enhancement factor. The measured photocurrent and nonlinearity after long-term operation were sufficient to continue using the device for harmonic mixing and waveform detection. Our measurements presented in the main text operated at ≥0.8 V/nm, which corresponds to a Keldysh parameter of $\gamma$ ≤0.6, indicating that we were operating in the tunneling regime.

Regarding device degradation, we note that previous work from Yang *et al.* ((Yang et al. 2020)) observed that modifying the antenna tips shifts its resonance and reduces peak field enhancement. This was attributed to reshaping due to heating from time-averaged power. In this study, we operated off-resonance and at lower repetition rates, significantly reducing average power density (by more than one order of magnitude). Consequently, we did not observe significant signal degradation over time after device "burn-in," as mentioned earlier.

## Degenerate Waveform Analysis

For degenerate waveform analysis of 0.177 PHz (1690 nm) as shown in SI. Fig. 9, we used an MZI to generate a pulse pair. We used ND filters to attenuate the gate and signal pulses and placed a chopper in the signal arm as a reference for the measurement of the current-induced change when the small signal perturbed the sub-optical cycle gate. After the pulse pair was recombined at the second beamsplitter, we used a linear polarizer with extinction coefficients greater than $10^5$ at 1690 nm.

We performed frequency-resolved optical gating (FROG) measurements of the pulses in time to provide a reference to compare against the measured fields. The error for the reconstruction shown is $5.510^{-3}$. In Fig. 10a, b we show the measured and retrieved spectrograms of the 10-cycle 1690 nm pulse used in the measurements. The degenerate sampled fields in Fig. 10c show relatively good agreement in the time domain with the FROG result. In both traces a small side lobe is present in time near 65 fs. The retrieved pulse duration was 58 fs FWHM and the sampled pulse duration was 57 fs FWHM. Lastly, the frequency domain comparison between the sampled intensity, retrieved intensity, and the grating-based spectrometer is shown in Fig. 10d. The sampled group delay and retrieved group delay show very good agreement.

We also performed degenerate waveform analysis using a 63-cycle pulse with a center frequency of 0.291 PHz (1030 nm) using an air-cooled 6-watt LightConversion Carbide after characterizing the pulse using SHG FROG. To begin our measurement, we employed the built-in pulse picker to reduce the repetition rate down to 500 kHz from 1 MHz. We then directed the output of the laser through the MZI with a closed-loop linear piezo stage delay line (Smaract SLC2445-S with MCS2) with $\pm$ 40 nm ($\pm$ 0.13 fs) repeatability. Utilizing a pristine device, we measured the current as a function of the gate pulse energy. We proceeded with measurements once the total photocurrent reached approximately 20 pA (blue line in Fig. 11a), which is equivalent to 1.4 nJ (0.328 V/nm) for this specific device measured. After measurements, we remeasured the current as a function of the gate pulse energy and observed degradation (orange curve Fig. 11a) associated with long pulse durations and highly localized fields at the nanoantenna tips. We tried to obtain 20 pA of current after measuring, however, this specific device rapidly degraded at 2.3 nJ. Using a signal pulse energy of 0.02 nJ (38.6 V/um), we degenerately sampled the pulse as shown in Fig. 11b.

In the top panel (1.76 ps x-axis range) of Fig. 11b, at $\pm$ 500 fs, small side lobes are seen. As we continue zooming in until the 10 fs x-axis range, the measured optical period is 3.4 fs and the expected optical period for a center frequency of 291 PHz is 3.4 fs. The technique can be generalized beyond 10-cycle pulses, however, since the devices were not operating in the tunneling regime for these measurements, these results have been placed in the supplemental information and require future studies to better understand the mechanism.

To verify the sampled field, we compared our measured results with SHG FROG. As seen in Fig. 11c, d is the measured and retrieved spectrograms of the 63-cycle 1030 nm pulse used in the measurements. The degenerate sampled fields in Fig. 11e show relatively good agreement in the time domain with the FROG result demonstrating that there is a small side lobe in time at $\pm 500$ fs. The retrieved pulse duration is 218 fs and the sampled pulse duration is 216 fs. Lastly, shown in Fig. 11f is the frequency domain comparison between the sampled intensity, retrieved intensity, and the grating-based spectrometer. The sampled group delay and retrieved group delay are also shown and show reasonable agreement.

## Non-degenerate Waveform Analysis

For non-degenerate waveform analysis of 0.353 PHz (850 nm) using a 0.177 PHz (1690 nm) gate, we modified the previously shown setup by adding a telescope for the 1690 nm gate to control the beam spot size before focusing on the nanoantennas. This modification was performed since when the SHG was generated and collimated using the type-1 1.5 mm thick BBO the beam spot was smaller in size before entering the objective. We also used a shortpass filter which was ND2 at wavelengths longer than 1450 nm as well as an achromatic half waveplate to rotate the SHG such that its polarization matched the gate pulse polarization. After the second beamsplitter, the pulse pair passed through a linear polarizer which had an extinction ratio that is $> 10^5$ at 1690 nm and $> 10^6$ at 850 nm.

We performed FROG measurements on the SHG, and the measured and retrieved spectrograms of the SHG of the 10-cycle 1690 nm pulse used in the measurements are shown in Fig. 13a, b. The error for the reconstruction shown is $1.310^{-2}$. The non-degenerate sampled fields in Fig. 13c show relatively good agreement in the time domain with the FROG result. The retrieved pulse duration was 49 fs FWHM and the sampled pulse duration was 48 fs FWHM. Lastly, shown in Fig. 13d is the frequency domain comparison between the sampled intensity, retrieved intensity, and the grating-based spectrometer. The sampled group delay and retrieved group delay also show very good agreement.

To demonstrate non-degenerate waveform analysis of a lower frequency using a higher frequency and to show that lower frequencies that are not integer harmonics can still be sampled, we used a separate few-cycle supercontinuum source with details that can be found in (Putnam et al. 2019). After supercontinuum generation, an SF10 prism pair was used to spatially disperse the beam. After the prism pair, the Er pump was spatially filtered, and the soliton and dispersive wave contributions were spatially separated. The higher-frequency dispersive wave portion of the supercontinuum was returned through the prism pair for compression and the removal of spatial chirp and was used as the gate. We note that the lower-frequency soliton, which was used as the signal, was not returned through the prism pair leaving some residual spatial chirp.

For the gate pulse, 19 pJ (0.21 V/nm at focus) of the 3-cycle (12 fs) dispersive wave (0.250 PHz, 1200 nm) was used. For the signal, 30 pJ (0.15 V/nm at focus) of the 11-cycle (65 fs) soliton (0.166 PHz, 1600 nm) was used. The measurement configuration is shown in detail in Fig. 14a. Degenerate waveform analysis of the dispersive wave was performed and is shown in Fig. 14b. Next, we performed non-degenerate waveform analysis measurement of the soliton and measured an optical period of approximately 6.0 fs as expected as Fig. 14c. Lastly, the normalized Fourier-transformed time-domain sampled waveforms are shown in Fig. 14d. This emphasizes the ability of the nonlinear mixing process to also sample lower frequency contributions at non-integer harmonics in the few-cycle regime without CEP locking.

## Regarding Alternative Materials

As these devices are based on resonant nanoantennas, we performed simulations using platinum, a metal that is not plasmonic, to demonstrate other nanoantenna materials could be used. It can be observed that toward DC frequencies (Fig. 15a), the field enhancement converges. It can be easily seen in Fig. 15b, where in the long wavelength field enhancement of the gold and platinum antennas converge, which is a characteristic of metal nanoantennas.

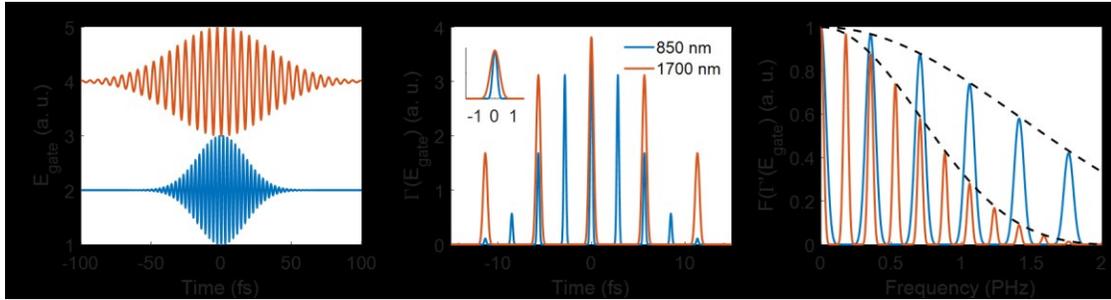

*Figure 5:* ***The transfer function response as a function of gate frequency.*** *We use a 10-cycle Gaussian pulse (a) with a carrier frequency of 0.353 PHz and 0.177 PHz to illustrate the frequency bandwidth dependence when the gate pulse changes. (b) The corresponding time domain electron burst for the two gate wavelengths used. (c) The frequency response of the electron burst for the two gate wavelengths used and their corresponding envelope if a 1-cycle pulse was used (dashed lines)*

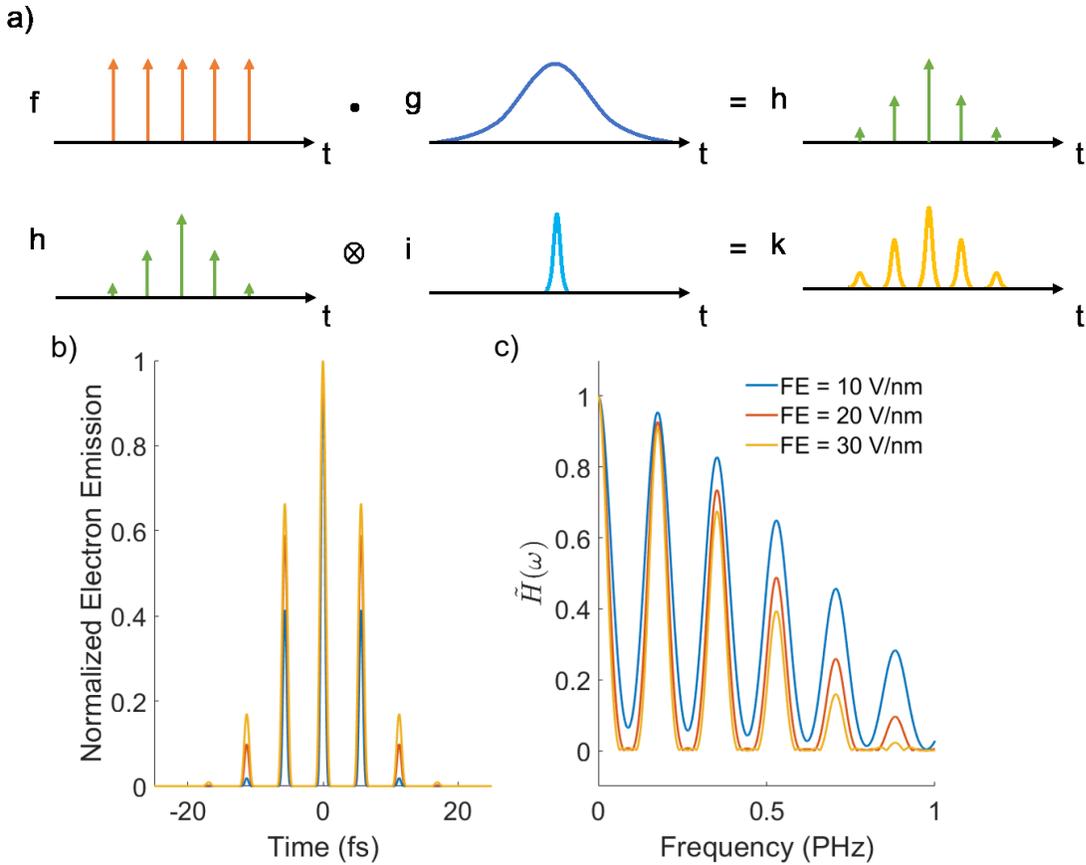

*Figure 6: **Illustration of comb generation in time and how the comb is affected by field enhancement.** (a) Illustration on the electronic comb generation in time. (b) The effect of the electronic comb with varied field enhancement (FE) using a 5-cycle $f_{\text{gate}}$ = 0.177 PHz optical field. The corresponding sub-optical-cycle electron emission burst is shown in (b) and the corresponding Fourier transform is shown in (c) as the field enhancement is increased.*

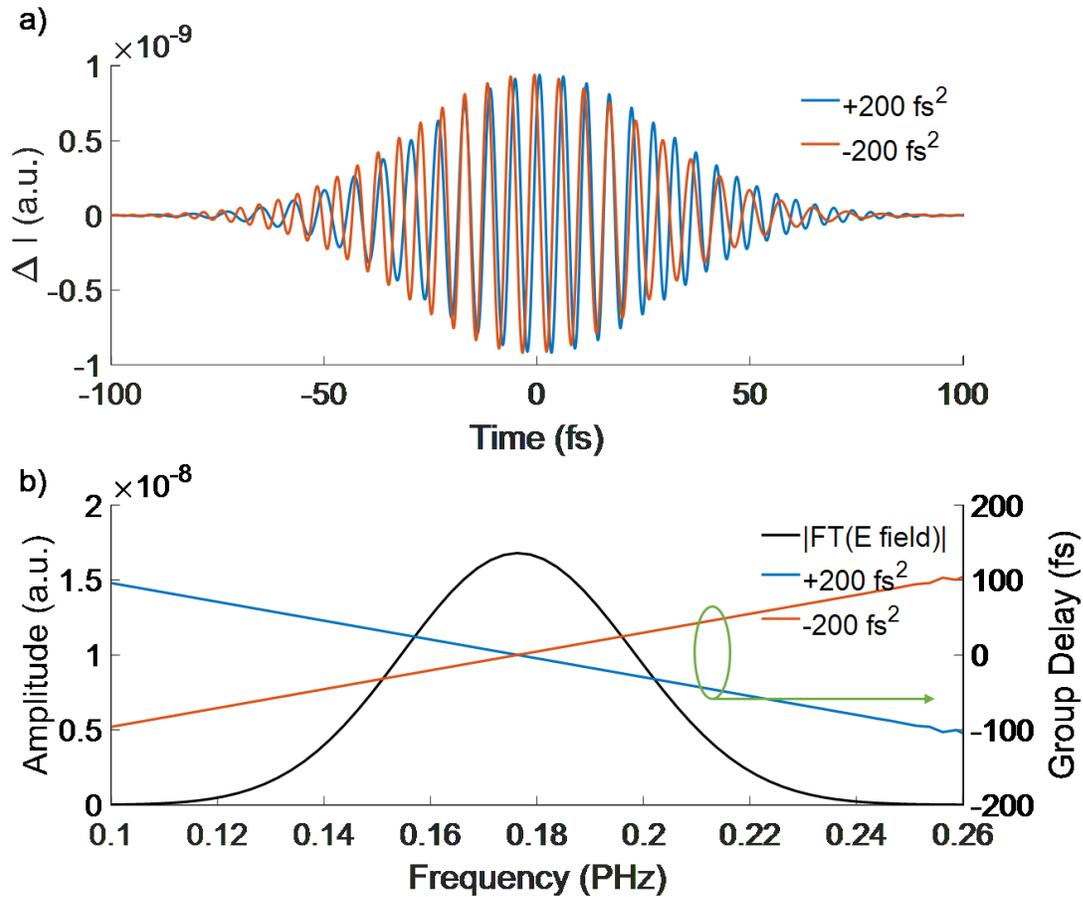

*Figure 7: **Simulated sampling with a non-zero GDD signal pulse.** (a) 2-cycle 1690 nm signal pulse with + 200 fs² (blue) and - 200 fs² (orange) GDD sampled using a 2-cycle 1690 nm gate pulse. (b) The corresponding Fourier-transformed sampled fields in (a) is shown as the black line while the group delays are shown in blue and orange.*

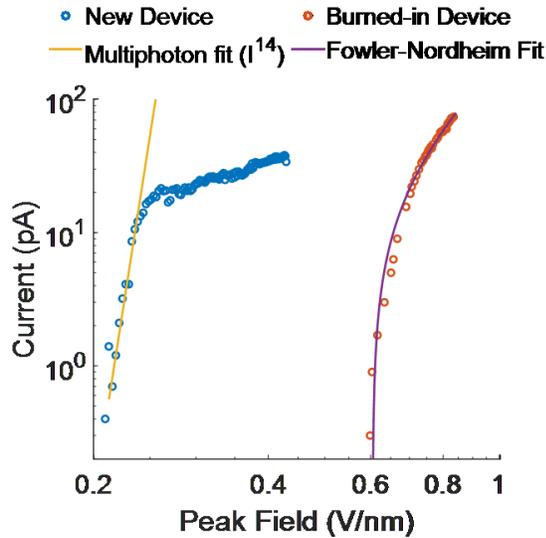

*Figure 8:* **Current output of the gold nanoantenna network.** *The curves include a new device and the same device after performing measurements for >4 hours using the 10-cycle f = 0.177 PHz.*

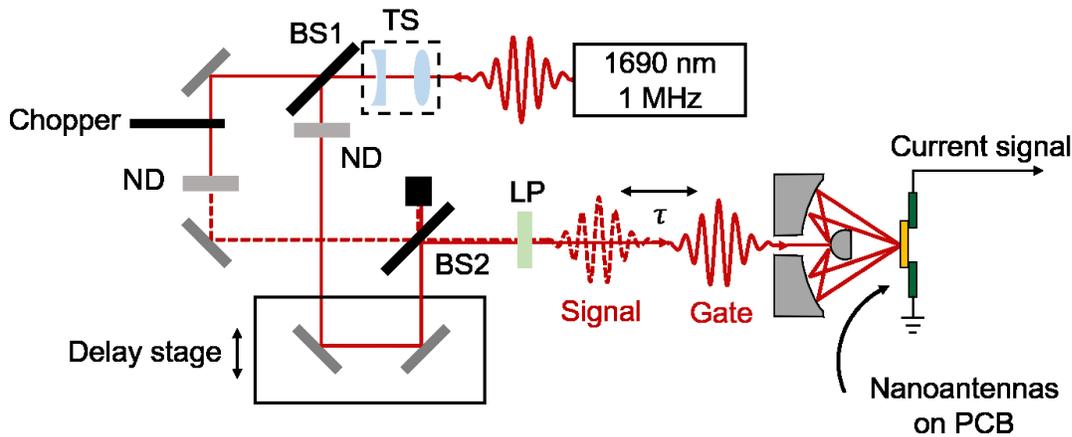

*Figure 9:* **A schematic of the experimental setup for the measurement of f = 0.177 PHz.** *The laser light was split using a beamsplitter (BS). One arm has a delay stage and was used as the gate pulse, while the signal arm was chopped and neutral density filters were used to attenuate the signal. Eventually, the two pulses are recombined using an identical beamsplitter before being sent to a reflective objective where they are focused onto the nanoantenna devices.*

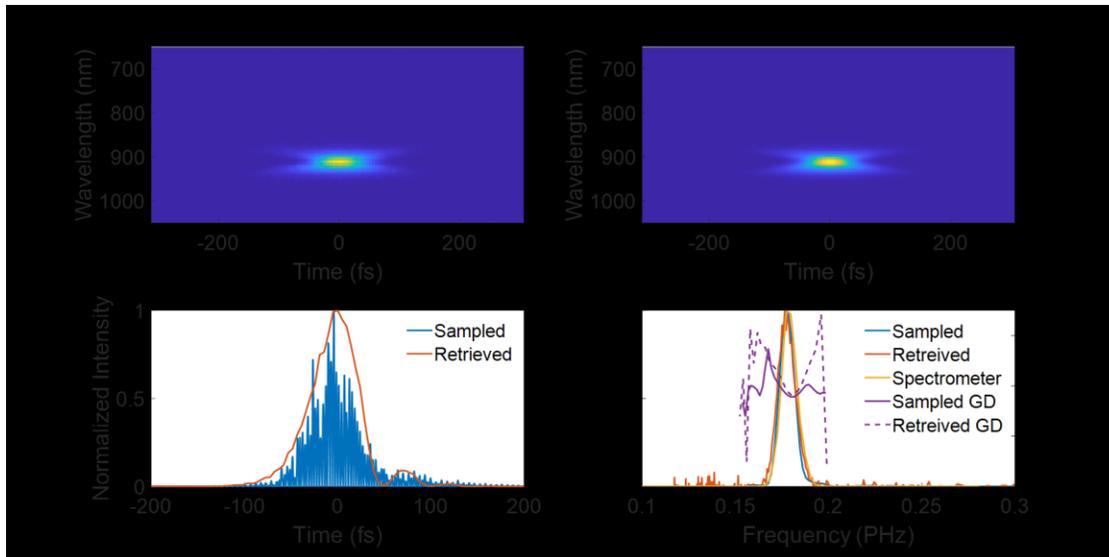

*Figure 10:* **Frequency-resolved optical gating (FROG) of the $f = 0.177$ PHz pulse.** *(a) measured and (b) retrieved spectrograms. (c) A comparison of the squared modulus of sampled optical fields and the retrieved pulse versus time. (d) A comparison of the squared modulus of the Fourier transformed sampled fields, retrieved pulse versus frequency, a spectrometer reference, group delay from the sampled optical fields, and the retrieved group delay.*

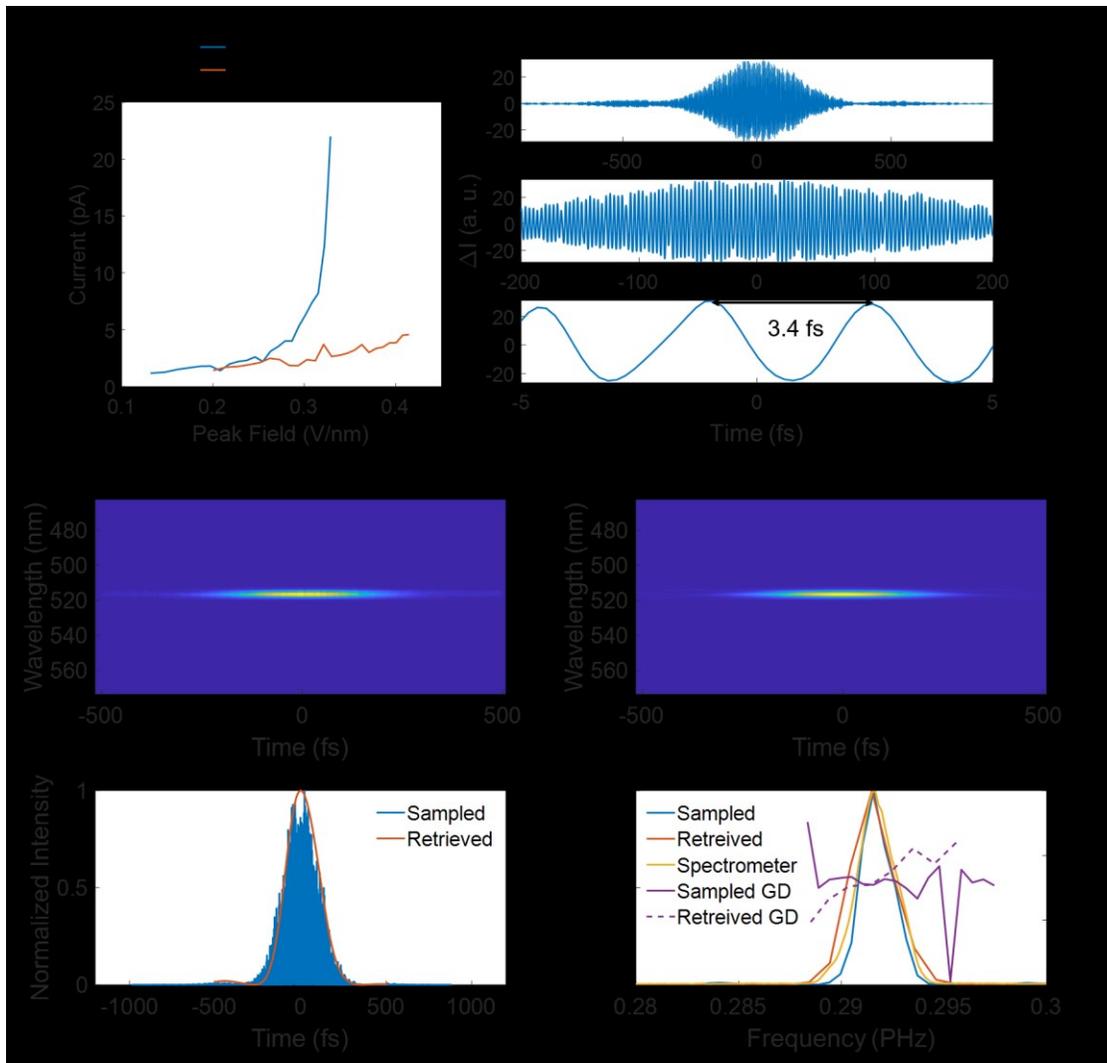

*Figure 11: **Degenerate optical field sampling of a 63-cycle (218 fs) 0.291 PHz pulse.** (a) The corresponding current vs peak field before and after sampling. (b) The degenerately sampled field with varying x-axis limits of 1760 fs, 400 fs, and 10 fs. The frequency-resolved optical gating (FROG) (c) measured and (d) retrieved spectrograms. (e) A comparison of the squared modulus of sampled optical fields and the retrieved pulse versus time. (f) A comparison of the squared modulus of the Fourier transformed sampled fields, retrieved pulse versus frequency, a spectrometer reference, group delay from the sampled optical fields, and the retrieved group delay.*

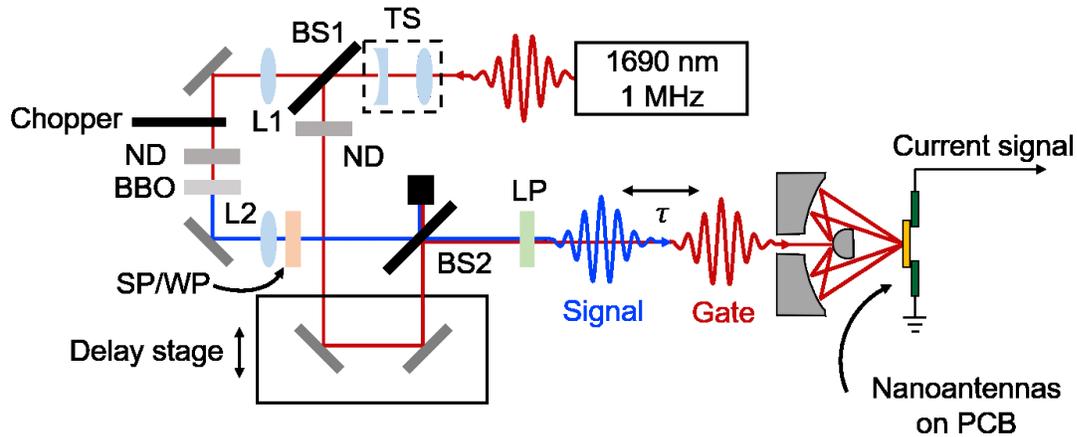

*Figure 12:* **A schematic of the experimental setup for the measurement of the $f_\text{gate}$ = 0.177 PHz and $f_\text{signal}$ = SHG of 0.177 PHz.** *The laser light first passes through a telescope and is split using a broadband beamsplitter (BS). One arm has a delay stage and was used as the gate pulse, while the chopped signal arm has a lens to focus onto the nonlinear crystal used to double the frequency and another lens was used to collimate the SHG. To control the SHG power, an ND filter was placed before the BBO and a 0.207 PHz (1450 nm) high-frequency pass filter (ND 2 at 0.177 PHz) was used to attenuate the residual $f$ = 0.177 PHz and a broadband achromatic waveplate (WP) was used to rotate the SHG polarization from vertical to horizontal. The two pulses are recombined using an identical broadband beamsplitter before being sent to a reflective objective where they are focused onto the nanoantenna devices*

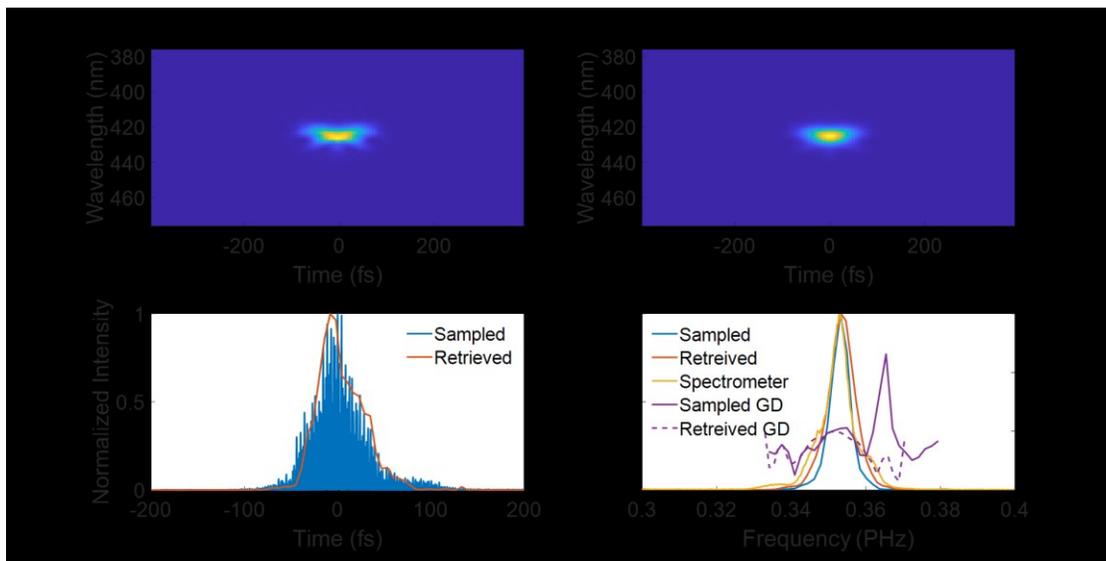

*Figure 13:* **Frequency-resolved optical gating (FROG) of the $f = 0.353$ PHz pulse.** *(a) measured and (b) retrieved spectrograms. (c) A comparison of the squared modulus of sampled optical fields and the retrieved pulse versus time. (d) A comparison of the squared modulus of the Fourier transformed sampled fields, retrieved pulse versus frequency, a*

spectrometer reference, group delay from the sampled optical fields, and the retrieved group delay.

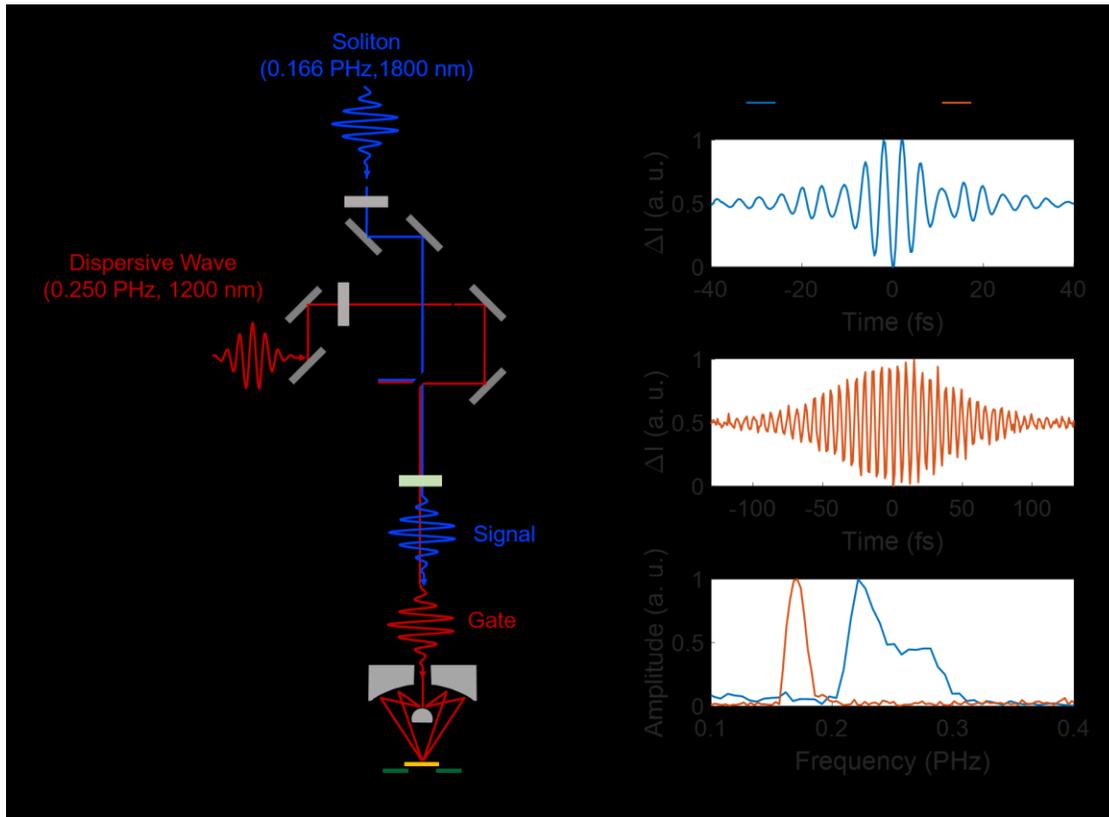

Figure 14: **Downsampling using a supercontinuum pumped by a 78.4 MHz Er Fiber Laser.** (a) The experimental schematic for the measurement. (b) Degenerate sampling of the dispersive wave. (c) Non-degenerate sampling of the soliton using the dispersive wave. (d) The corresponding Fourier transformed sampled fields.

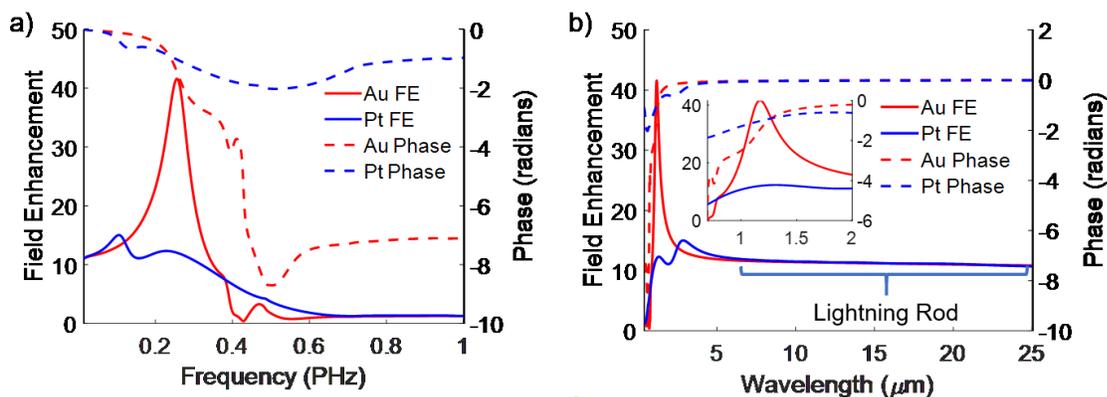

Figure 15: **FDTD simulation of the gold nanoantenna and platinum nanoantenna.** The platinum serves as a reference for a non-plasmonic metal. In (a) is the field enhancement and phase as a function of frequency while (b) is in wavelength

## Data and code

We have provided the original data and selected code used in this manuscript.